\documentclass[conference]{IEEEtran}

\usepackage{graphicx} 
\usepackage{booktabs} 
\usepackage{amsmath}
\usepackage{amsfonts}
\usepackage{amssymb}
\usepackage{epstopdf}
\usepackage{dblfloatfix}
\usepackage[caption=false,font=footnotesize]{subfig}
\usepackage{color}
\usepackage{enumitem}
\usepackage{amsthm}
\usepackage{algorithmic}
\usepackage{multirow}
\usepackage{array}
\usepackage{cite}
\begin{document}

\title{Distributed Proportional-Fairness Control in MicroGrids via Blockchain Smart Contracts}

\author{\IEEEauthorblockN{Pietro Danzi, Marko Angjelichinoski, \v{C}edomir Stefanovi\'c, Petar Popovski}
\IEEEauthorblockA{Department of Electronic Systems, Aalborg University, Denmark \\
Email: \{pid,maa,cs,petarp\}@es.aau.dk }
}

\maketitle

\begin{abstract}

Residential microgrids (MGs) may host a large number of Distributed Energy Resources (DERs).
The strategy that maximizes the revenue for each individual DER is the one in which the DER operates at capacity, injecting all available power into the grid.
However, when the DER penetration is high and the consumption low, this strategy may lead to power surplus that causes voltage increase over recommended limits.
In order to create incentives for the DER to operate below capacity, we propose a proportional-fairness control strategy in which (i) a subset of DERs decrease their own power output, sacrificing the individual revenue, and (ii) the DERs in the subset are dynamically selected based on the record of their control history.
The trustworthy implementation of the scheme is carried out through a custom-designed blockchain mechanism that maintains a distributed database trusted by all DERs.
In particular, the blockchain is used to stipulate and store a \emph{smart contract} that enforces proportional fairness.
The simulation results verify the potential of the proposed framework. 
\end{abstract}

\IEEEpeerreviewmaketitle

\section{Introduction}

The deployment of Distributed Energy Resources (DERs) in the residential low voltage (LV) microgrids (MGs) aims to improve their self-sustainability and reduce the transmission losses \cite{ref1m}.
DERs based on renewable resources, such as solar photovoltaic (PV), are traditionally operated at capacity to inject all available power in the grid and thus maximize the efficiency, regardless of the grid state.
However, the variability of the capacity may cause erratic voltage behavior on the distribution feeders.
In particular, grids with high penetration of PVs may experience voltage increase over the recommended levels, e.g., during afternoons, when the production is high and the household consumption is low \cite{pedersen2015disc}.
An approach to prevent grid instability due to dramatic voltage increases is to control the power output of DERs, e.g., via active power curtailment or reactive power adjustment \cite{tonkoski2011coordinated}. 
In this respect, \cite{pedersen2016fairness} proposes a control strategy based on the principle of proportional fairness, where all DERs equally contribute to voltage regulation all the time; the strategy is executed by the remote central authority such as the distribution system operator (DSO).
However, enforcing that all DERs participate in voltage control and curtail their output power all the time is characterized with a control complexity that increases with the number of DERs and a reduced possibility for the owners to operate an economic strategy.
Moreover, existence of the centralized authority involves the issue of Single Point of Failure (SPoF).

In this paper, we propose a novel control scheme based on the proportional fairness in which (i) only a subset of DERs act as voltage regulators and curtail their individual power outputs over control periods, and (ii) ensures that, in the long term, DERs participation in voltage regulation is balanced.
To give the DERs the incentive to fairly participate in voltage regulation, we introduce a principle based on exchange of \emph{credits}. 
Specifically, in order to join the regulating subset, DER asks for a credit, which will be paid by the DERs that are not in the regulating subset and therefore operate at their full capacity. 
In turn, the decrease of credit status of the DERs that have not participated in voltage regulation ultimately forces them to participate in the voltage regulation in future.

In order to avoid existence of a central authority and, thus, a SPoF, the proposed protocol runs in a distributed manner.
The credit statuses of all DERs are tracked by the use of a blockchain protocol, initially introduced with Bitcoin cryptocurrency \cite{nakamoto2008bitcoin}.
Its major advantage is the capability to implement a distributed database that serves as record of the system's state and of its history, trusted by all agents, in this case DERs.
All agents store identical copies of the database.
The database is hard to tamper with, since an agent can add new record to it \emph{only} if a proof-of-work (POW) is obtained, where POW is the solution of a computational puzzle that requires an investment of electrical energy to run the computation.
In particular, the blockchain is used to memorize (i) the state of \emph{smart contracts} \cite{wood2014ethereum}, which are computer programs that can receive, store and pay credit, i.e., cryptocurrency, and (ii) the credit history of agents.
In the proposed framework, a smart contract acts as a trustworthy \emph{distrubuted} control authority, realized via a custom-designed blockchain mechanism operated by all DERs.
Specifically, DERs that are installed on the same distribution feeder stipulate a smart contract among them, determining which units will act as voltage regulator over control periods, based on their available credit statuses and the economic strategy that they individually adopt.
In the long term, this ensures a fair rotation in the participation of the DERs to the MG regulation. 

The rest of the paper is organized as follows.
Section~\ref{sec:problem_formulation} introduces the system model and the proportional-fairness control strategy assuming a centralized setup.
Section~\ref{sec:blockchain} introduces the main concepts of a blockchain protocol.
Section~\ref{sec:solution} presents the distributed, blockchain-based architecture fostering the proportional-fairness control.
Section~\ref{sec:case_study} contains a case study based on a simulation of a power system and an instance of blockchain protocol, verifying the proposed framework.
Section~\ref{sec:conclusion} concludes this paper.

\section{System Model}
\label{sec:problem_formulation}

\begin{figure}[!tb]
\centering
  \includegraphics[width=\columnwidth]{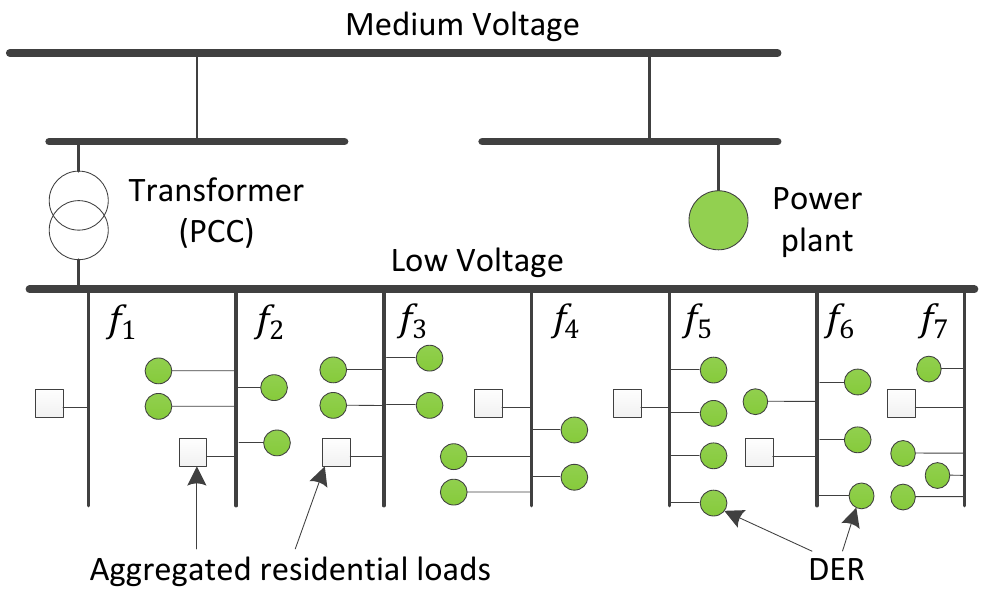}
    \caption{The system model.}
    \label{fig:grid}
\end{figure}

We consider a LV distribution grid (LVDG) composed of $N_{\text{MG}}$ alternate-current MGs.
A MG~$f$ hosts $U_f$ DERs, $U_f \geq 0$, which are PV generators that supply residential loads, see Fig.~\ref{fig:grid}.
The total number of DERs in the LVDG is $U_{\text{total}} = \sum_{f=1}^{N_{\text{MG}}} U_f$.
All MGs jointly strive to maintain the LVDG grid voltage amplitude within acceptable limits \cite{rocabert2012control}.
To do so, in each MG there is a dedicated \emph{voltage regulator} elected from the local DERs. 
In the rest of this section, we first describe the voltage regulation mechanism.
Afterwards, we introduce the regulator election strategy that each MG employs. 

\subsection{Voltage regulation}
 
Assume MG~$f$ with $U_f$ DERs indexed in the set $\mathcal{U} = \{1,2,...,U_f\}$.
Each DER is connected to the MG via a power electronic converter (PEC) that controls its output and that supports dual mode capability \cite{ref12}, i.e., a PEC can operate in current source converter (CSC) or voltage source converter (VSC) mode.
In CSC mode, DER~$u$ is operated at capacity, outputting all available power $g_u$ using maximum power point tracking algorithm, and is \emph{not} capable of regulating the voltage \cite{rocabert2012control}.
In VSC mode, DER~$u$ acts as the voltage regulator at the expense of the reduced active output power $p_u$, $p_u \leq g_u$.
The voltage is regulated via the active output power using the following droop control law \cite{guerrero2011hierarchical}:
\begin{equation}\label{droop}
 v_u = v_{ref} - \gamma(g_u - p_{u}),
\end{equation}
where $v_u$ and $v_{ref}$ are the voltage at the output of the DER and the reference voltage of the MG, respectively, while $\gamma$ is the droop parameter, chosen such that (i) $v_u$ is maintained within tolerable limits $v_{\min}$ and $v_{\max}$, and (ii) all VSCs in the LVDG achieve proportional power sharing.
The output reactive power $q_u$ is determined via the active output power, subject to a constraint on the apparent power, as follows:
\begin{equation}
q_u = \sqrt{s_{\max}^2 - p_u^2},
\end{equation}
where $s_{\max}^2$ is the maximum tolerable apparent power.\footnote{We note that more sophisticated control schemes, based on both active and reactive power adjustments, can be adopted for the voltage regulation. However, they are beyond the scope of the paper.}
The droop control law \eqref{droop} clearly shows that, in VSC mode, the active power of DER $u$ is below capacity, leading to revenue loss of the DER owner.
Nevertheless, the presence of VSC units is an imperative for voltage regulation of the MG.

\subsection{VSC election per MG based on proportional fairness}
\label{sec:VSC}

\begin{figure}[!tb]
\centering
  \includegraphics[width=0.85\columnwidth]{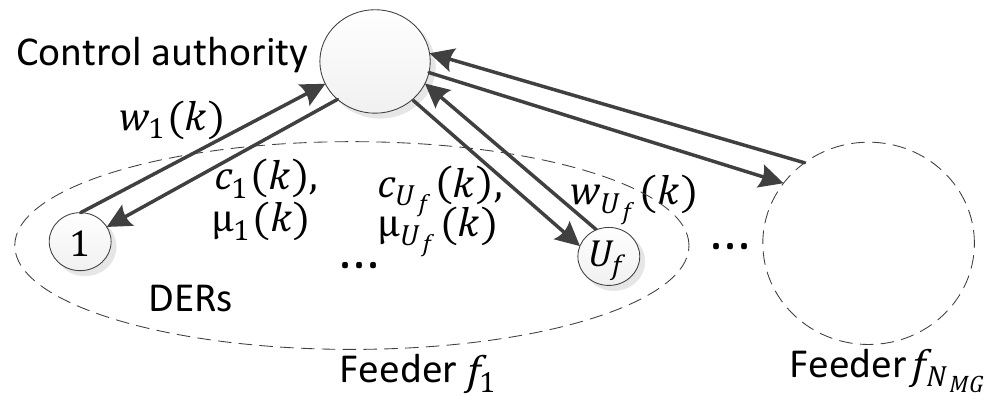}
    \caption{Message exchanges between the DERs and the control authority.}
    \label{fig:architecture}
\end{figure}

We first formulate the centralized version of the VSC election strategy, employed in each MG, and enabled by an external control authority, e.g., the DSO, see Fig.~\ref{fig:architecture}.
The strategy runs periodically every $T^{tc}$ seconds, where the interval between two consecutive VSC assignments is referred to as \emph{control period}, and is sufficiently long to allow reliable exchange of messages among DERs and the authority.
Let $\mu_u (k)$ denote the operating mode of DER $u$ in arbitrary control period $k$.
Hence, $\mu_u (k)$ is a binary variable:
\begin{equation}
\mu_{u}(k) =
\begin{cases}
0 \quad \text{if DER }u \text{ operates as CSC in period }k, \\
1 \quad \text{if DER }u \text{ operates as VSC in period }k.
\end{cases}
\end{equation}

Denote by $\boldsymbol{\mu}_u$ the sequence of operating modes of DER $u$ in all periods up to interval $K$, i.e., $\boldsymbol{\mu}_{u} = [ \mu_{u}(0), \hdots, \mu_{u} (K-1)]$.
Stacking $\boldsymbol{\mu}_u,~\forall u\in\mathcal{U}$, vertically, we form the $U_f\times K$ matrix of control mode histories of all DERs in the MG up to period $K$, denoted by $\mathbf{M}$, where $[\mathbf{M}]_{u,k} = \mu_{u}(k)$.
The $k$-th column of $\mathbf{M}$ represents the operating modes of all DERs in the MG in control period $k$, and is denoted by {$\boldsymbol{\mu}^{(k)}$}. 
The objective of the VSC election strategy is to ensure fairness among DERs in the MG, i.e., all DERs should equally participate in voltage control over time.
This can be expressed as follows:
\begin{equation}\label{eq:objective}
\mathbf{M} \, \mathbf{1}_{K} = \frac{K}{U_f} \mathbf{1}_{U_f},
\end{equation}
where $\mathbf{1}_{U_f}$ is the all-ones vector of length $U_f$.
In addition, the strategy should also ensure that only one DER per MG acts as VSC in each control period, i.e.:
\begin{equation}
\mathbf{1}_{U_f}^{\mathsf{T}} \, \boldsymbol{\mu}^{(k)} = 1,
\end{equation}
with $(\cdot)^{\mathsf{T}}$ denoting the transpose operator.

\begin{figure*}[!tb]
\centering
\subfloat[]{\includegraphics[width=0.45\columnwidth]{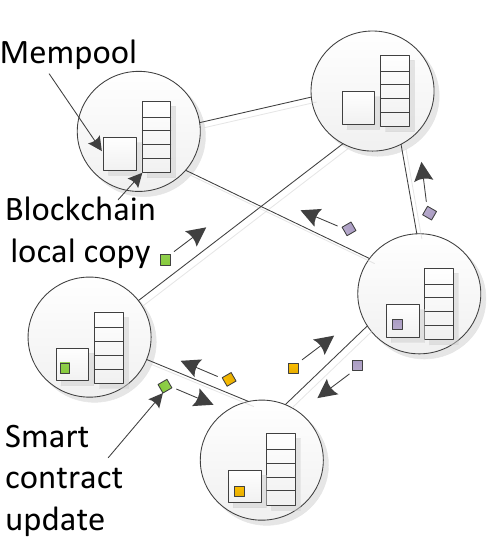}}
\hfill
\subfloat[]{\includegraphics[width=0.45\columnwidth]{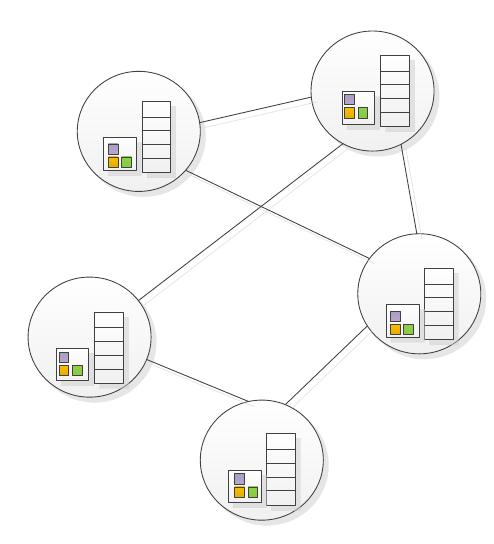}}
\hfill
\subfloat[]{\includegraphics[width=0.45\columnwidth]{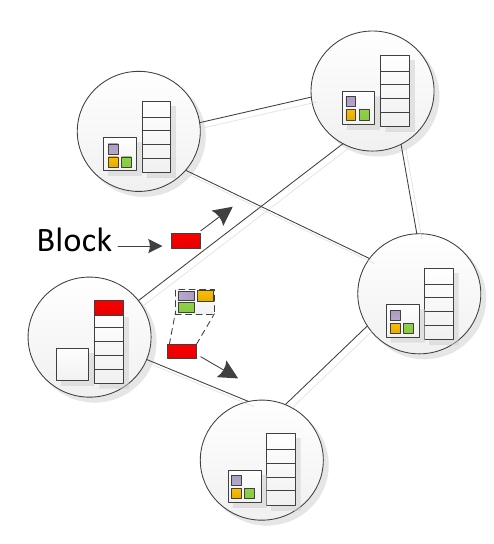}}
\hfill
\subfloat[]{\includegraphics[width=0.45\columnwidth]{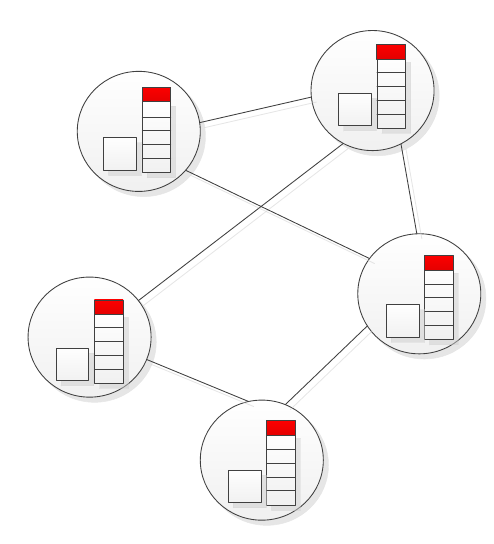}}
\caption{Different phases of blockchain protocol: (a) contract updates are propagated in the network, (b) contract updates are accumulated in the miners' buffers, (c) a new block (red packet) that includes the updates is generated and propagated and (d) the new block is accepted by the entire network.}
\label{fig:blockchain}
\end{figure*}

DERs are given incentive to participate in voltage regulation through acquisition of \emph{credit} when they operate in VSC mode.
Denote by $c_u(k)$ the credit status of DER $u$ in the control period $k$ and let $\mathbf{c}^{(k)} = [c_1(k),\hdots,c_{U_f}(k)]$. 
In the proposed scheme, the total available credit is constant, i.e., $\sum_{u=1}^{U_f} c_u (k) = C$, $\forall k$, and the credit is only redistributed among DERs, ultimately forcing them to act as VSCs.

The redistribution of the credit is performed in the following way. 
In period $k$, DER $u$ sends to the control authority the credit demand $w_u(k)$ asked for performing the role of VSC for the MG in the next period $k+1$.
The credit demand is chosen in the interval $[0,c_u(k)]$, according to the individual strategy adopted by DER $u$.
For the sake of simplicity, in the rest of the paper we assume that DER $u \in \mathcal{U} $ chooses $w_u(k)$, $\forall k$, randomly using uniform distribution:
\begin{equation}\label{eq:constrain}
w_u(k) \sim \text{unif} \left( 0,c_u(k) \right).
\end{equation}
The control authority stores the information sent by all DERs in the MG in the vector $\mathbf{w}^{(k)} = [w_1(k),\hdots,w_{U_f}(k)]$, and chooses the DER with the \emph{lowest} credit demand to operate as VSC in the next control period:
\begin{equation}\label{eq:decision}
\hat{u} = \text{argmin}_{u} \mathbf{w}^{(k)}.
\end{equation}
Then, the control authority sends the messages about (i) the control modes of all the DERs in the MG in the next control period $\boldsymbol{\mu}^{(k+1)} = \mathbf{e}_{\hat{u}}$, where $\mathbf{e}_{\hat{u}}$ is a vector with $1$ at position $\hat{u}$ and $0$ elsewhere, and (ii) the update of the total credit
\begin{equation}\label{eq:credit}
\mathbf{c}^{(k+1)} = \mathbf{c}^{(k)} - \mathbf{w}^{(k)} + \left( \mathbf{1}_{U_f}^{\mathsf{T}}\mathbf{w}^{(k)} \right) \mathbf{e}_{\hat{u}},
\end{equation}
see Fig.~\ref{fig:architecture}.
A careful inspection of \eqref{eq:decision} shows that the total credit is redistributed by (i) taking from the DERs their respective credit demands and (ii) giving the total demanded credit to the DER $\hat{u}$ that had the lowest demand and that will take the VSC role in the next period.

The rule for choosing the credit demand \eqref{eq:constrain} and the VSC election rule \eqref{eq:decision} show that DERs with lower credit status have higher chances of operating as VSC.
On the other hand, low credit status implies that the DER has previously infrequently operated as VSC.
In this way, the proportional fairness is promoted among DERs in the MG.
We also note that the results provided in Section~\ref{sec:case_study} verify that the objective \eqref{eq:objective} becomes satisfied as the number of control period increases.

To run the VSC election strategy, the LVDG requires a communication network to interconnect the DERs with the central authority to support the message exchange depicted in Fig.~\ref{fig:architecture}. 
This can be implemented using wireless (e.g., cellular) network, or wired network (e.g., power line communications).

Finally, the election strategy \eqref{eq:decision} and the credit update rule \eqref{eq:credit} require presence of the central authority's database that records the current credit status $\mathbf{c}^{(k)}$ and that is trusted by all agents. 
This need for the establishment of a trustful relation among agents prevents the implementation of simple decentralized control systems, such as token rings, in which DERs hold the control in turns. 
Motivated by this insight, we develop a decentralized control solution based on blockchain protocol, which establishes a trustful distributed record both of the credit status of all DERs and of the control history.

\section{Blockchain Protocol}\label{sec:blockchain}

In this section, we describe a version of the blockchain protocol that realizes only the functionalities required to support the proposed control strategy.
An introduction to the general variant of the blockchain can be found in \cite{narayanan2016bitcoin}.

A blockchain is a distributed database consisting of identical copies that are stored in the memory of each agent.
It is organized as a concatenated list of blocks that can be expanded by any agent with new blocks. 
Each block stores a set of smart contracts updates.
A smart contract is a computer program that can be executed by an agent that has the blockchain software.
As the agent executes the program locally, whenever it modifies the contract internal state, e.g., by exchanging credit with it, the rest of the agents should be informed in order to run the new version of the contract.
The new state is communicated through a smart contract update.
To avoid proliferation of different states of the contract, the blockchain in parallel executes the update verification process via \emph{block generation}, ensuring that only verified updates are included in the blockchain.
We proceed by providing the details.

Fig.~\ref{fig:blockchain} illustrates the blockchain operation via an example of a peer-to-peer network of interconnected agents.
Every agent is provided with a local copy of the blockchain, and a buffer, named ``mempool''.
When smart contract updates are produced, they are sent to the neighbors, Fig.~\ref{fig:blockchain}(a).
The updates are temporary stored in mempools, Fig.~\ref{fig:blockchain}(b), being not considered valid yet.
In parallel, agents are also generating blocks, i.e., working to solve the computational puzzle (by the blockchain protocol) and obtain POW. 
When an agent obtains POW and generates a new block, it fills it with the smart contract updates present in its mempool and transmits the new block to its peers, see Fig.~\ref{fig:blockchain}(c).
Upon the reception of the block, a neighbouring agent verifies that it contains a valid POW, by checking the provided solution of the puzzle.
If the verification succeeds, the agent adds the block to its blockchain, Fig.~\ref{fig:blockchain}(d), and propagates it further to its neighbors.
The verification and propagation of successfully verified block continues until all agents in the network are reached.
Finally, when all agents have received and verified the block, they have the same updated version of the blockchain, see Fig.~\ref{fig:blockchain}(d), and thus, their copies of the smart contract have the same internal state.
The agents also remove from their mempools the smart contract updates included in the received and verified blocks.

The smart contract can be abstracted as a \emph{virtual agent} that interacts with the actual agents according to the logic defined in its program, and in this way regulating credit redistribution among the actual agents through the contract updates.
In the context of the proposed framework, the smart contract updates are produced by the MG control application, while the block creation process is intrinsic to the blockchain and in charge of keeping it consistent.
These two process run independently, where the latter ensures that the smart contract updates are eventually stored in the blockchain. 

\subsection{Blockchain consistency}

As depicted in Fig.~\ref{fig:blockchain}(c)--(d), the blockchain consistency is guaranteed through propagation of newly generated blocks to all other agents.
In order to avoid the generation of uncontrolled amounts of blocks, a block generation requires an investment of resources.
In the seminal paper \cite{nakamoto2008bitcoin}, the investment is the consumption of electrical power required to solve a computational puzzle, known as POW. 
This process of obtaining POW is named \emph{mining}; the agent that first solves the POW decides the content of the next block.
The probability $p_b$ to generate a new block before the other agents depends on the computational resources allocated to solve the puzzle.
The difficulty of the puzzle is tuned to keep the block generation rate constant over time, see \cite{nakamoto2008bitcoin} for details.

It may happen that two or more new valid blocks are created concurrently by different agents, propagating in different subgraphs of the network.
In this case, the network is split over different, but valid versions of the blockchain.
The contention rule adopted by Bitcoin is ``the longest chain wins", i.e., the split is solved when some ``newer'' valid block is generated, received and accepted by all agents, updating the blockchain and making it consistent again \cite{narayanan2016bitcoin}.

\subsection{The cost of mining}\label{sec:fees}

In standard blockchain realizations, the resource investment in mining operation is compensated by attributing some credit, i.e., cryptocurrency, to the miners, either by (i) increasing the credit of the miner that generates the block, incrementing the total amount of credit in the system, or (ii) letting the miner demand a credit fee for each smart contract update that it includes in a block.
This stimulates miners to consume electrical power to obtain POW, and thus reduces the risk that a malicious miner may impose his version of the blockchain. 

In this work, the miners are the DERs in the MG, i.e., the blockchain is \emph{private}, but they are not rewarded with credit for mining.
Nevertheless, a DER is stimulated to mine, in order to ensure that the blockchain is consistent with its view of the credit status and control history.
Also, not providing credit for mining eliminates the possibility of DERs being more motivated in mining than in voltage regulation.

\subsection{Peer-to-peer network and protocol messages}

The blockchain is based on peer-to-peer networking, where each agent connects to a set of randomly selected neighbors.\footnote{It is required that the communication graph formed in this way is connected, but the related details are beyond the scope of the paper.}
The communication protocol adopts TCP connections to exchange two types of information, as shown in Fig.~\ref{fig:blockchain}.
\begin{enumerate}
\item The smart contract updates, Fig.~\ref{fig:blockchain}(a).
\item The blocks, Fig.~\ref{fig:blockchain}(c).
\end{enumerate}



\subsection{Security issues of blockchain protocol}\label{sec:sec_issue}

The blockchain protocol is founded on the assumption that the resources required to generate blocks are well distributed among agents, making it hard for a single agent to consecutively generate blocks and and impose its view of the blockchain that may potentially include false data \cite{narayanan2016bitcoin}.
This assumption can be adopted in the scenario considered in the paper, as it is expected that microcontrollers of the DERs have similar hardware characteristics.

Security threats to blockchain may also come from peer-to-peer networking.
A prominent example is the eclipse attack \cite{heilman2015eclipse}, in which all neighbors of an agent are under the control of a malicious agent.
In this case, the copy of the blockchain of the attacked agent may be compromised.
To avoid such scenario, the communication peers are selected randomly.


\section{The Proposed Distributed Control Framework}\label{sec:solution}

The proposed solution is enabled by a blockchain protocol (i.e., software) implemented by the agents, i.e., DERs of the distribution grid.
The mining process is done by agents themselves, where the agents do not cooperate and the POW is of low difficulty determined by their limited computational capabilities.
The agents have access to a peer-to-peer network, where the access is granted only to legitimate agents.
We assume that the rate at which POWs are obtained and new blocks propagated through the network is sufficiently high to prevent overflow of the mempools.
Finally, the message propagation delay is assumed negligible compared to the control period duration $T^{tc}$.

\begin{figure}[!tb]    
\centering
\subfloat[Credit sending.]{\includegraphics[width=0.39\columnwidth]{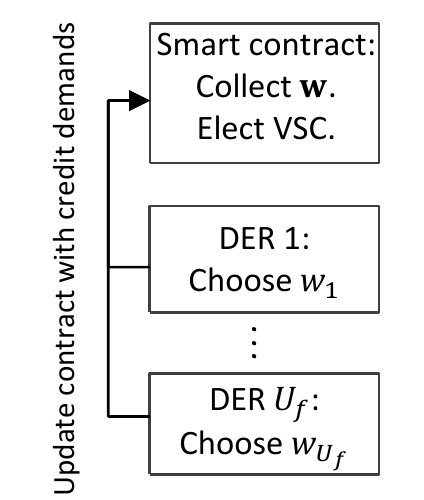}}
\hspace{2pt}
\subfloat[Credit receiving.]{\includegraphics[width=0.39\columnwidth]{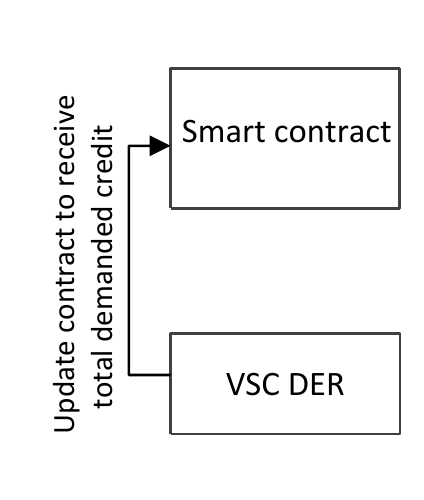}}
    \caption{Relationship among DERs in one feeder and their smart contract.}
    \label{fig:relationship}
\end{figure}

A smart contract is deployed for each grid feeder, playing the role of the central authority (i.e., DSO), resulting in a unique blockchain that stores $N_{MG}$ contracts. 
The relationship of DERs in a feeder and their smart contract is depicted in Fig.~\ref{fig:relationship}.
During a control period, each DERs chooses its credit demand for the next period via \eqref{eq:constrain}, and updates its local copy of the smart contract by \emph{sending}, i.e., transferring, the demanded amount to it, see Fig.~\ref{fig:relationship}(a).
These updates are propagated to peers, Fig.~\ref{fig:blockchain}(a)--(b), and the block generation process, Fig.~\ref{fig:blockchain}(c), ensures that all the copies of the smart contract have the consistent knowledge of them, Fig.~\ref{fig:blockchain}(d).
The credit sending is disabled during the final part of the period via the contract locking, to ensure the consistency of local copies at the control actuation instant, i.e., at the beginning of the new period.
At this point, based on its internal state, the smart contract (i.e., each its copy) elects the (same) VSC for the next control period using \eqref{eq:decision}, and DERs are locally notified (e.g., via reading the state of the smart contract) about their operating mode for the next control period.
In the next round, the elected VSC updates its local smart contract to receive the total demanded credit, see Fig.~\ref{fig:relationship}(b).
This triggers a new round of the update propagation, Fig.~\ref{fig:blockchain}(a)--(b), which is eventually stored in the blockchain after a block embedding it becomes generated and propagated through the network, Fig.~\ref{fig:blockchain}(c)--(d). 

\begin{figure}[!tb]

\centering
\subfloat[Blockchain-based scheme: the messages are exchanged among DERs.]{\includegraphics[width=\columnwidth]{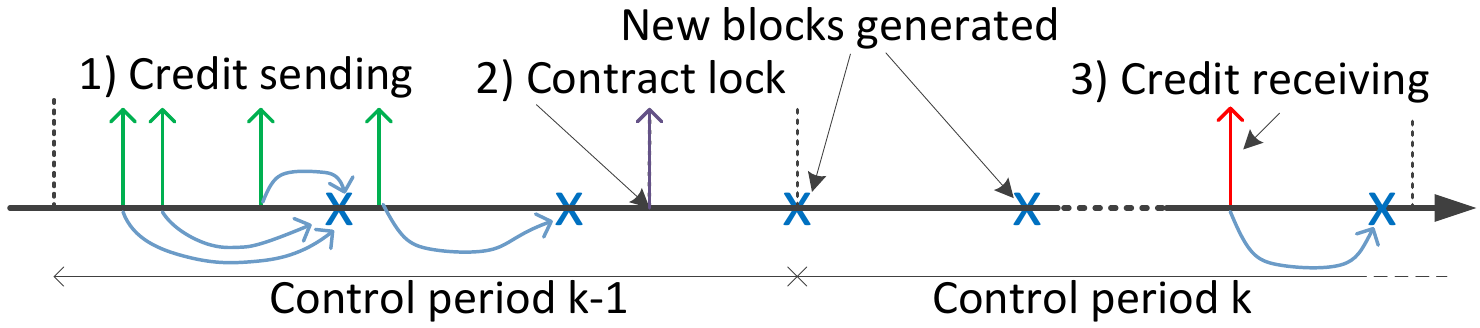}}
\vspace{6pt}
\subfloat[Centralized scheme: The messages are sent from DERs to the central authority (uplink) and from the central authority to DERs (downlink).]{\includegraphics[width=\columnwidth]{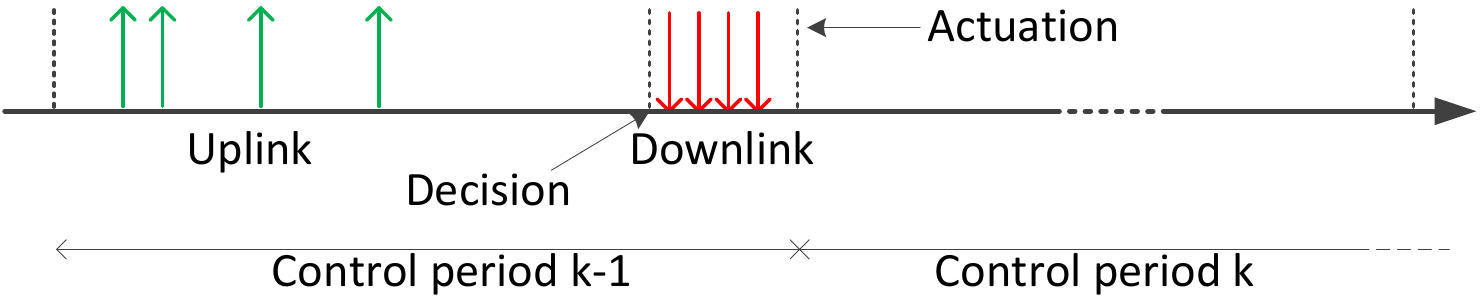}}

\caption{Sequence of the communication exchanges (a) in the blockchain-based and (b) in the centralized scheme.}
    \label{fig:comparison}
\end{figure}

For the sake of completeness, in Fig.~\ref{fig:comparison}(a) we expose the operating sequence of the framework related to a single control period, i.e., period $k$, assuming that the contract updates related to period $k$ start in period $k-1$:\footnote{In general, the framework can be operated such that the contract updates for period $k$ are started in a period $k_1$, where $k_1 \leq k - 1$, allowing to solve the eventual blockchain inconsistencies due to propagation and mining delays. The related analysis is beyond the paper scope.}
\begin{enumerate}
\item \emph{Credit sending:} During period $k-1$, each DER transfers to the smart contract the credit demanded to operate as VSC in the next period (represented by green arrows). These updates are propagated through the network and stored in the mempools, and gradually included in the blockchain when new blocks are generated. 
\item \emph{Contract lock and VSC election:}
At a predefined instant before the end of period $k-1$, all DERs modify the state of the contract to lock it (purple arrow).\footnote{Note that the contract is effectively locked only by the first contract update that is included in the blockchain, which invalidates the consecutive locks.} The VSC for control period $k$ is elected via \eqref{eq:decision}, uniquely over all copies of the smart contract. All DERs set their control mode accordingly.
\item \emph{Credit receiving:} The VSC DER withdraws the credit obtained for the control period $k$ from the smart contract (represented by the red arrow). To do this, it modifies the state of the contract, communicates the state update to the other agents, and waits for it to be included in a newly mined block. The credit stored by the contract can only be withdrawn by the DER operating as VSC.\footnote{All contract updates are certified with public-key cryptography, cf. \cite{nakamoto2008bitcoin}.}
\end{enumerate}
The figure also depicts the block generation process (represented by blue crosses) that is decoupled from the process of contract updates.
It may happen that some newly generated blocks do not contain any updates and such blocks only verify the consistency of the current state of the blockchain.

In the rest of this section, we outline several important aspects of the framework.

\subsection{Plug-and-play feature} 

The blockchain-enabled solution provides a smooth plug-and-play since the joining DER just needs to start interacting with the contract.
Observe that it is likely that the joining DER will be chosen as VSC in the next control periods, as its credit status is zero.

\subsection{Control availability and trustworthiness}

The decentralized solution does not suffer from the SPoF, providing an increased control availability with respect to the centralized architecture.
It also avoids reliance on data received by an external authority, which can be tampered without being detected by the MG agents.
On the other hand, the major weakness of the blockchain-based architecture is the modification of the database operated by a subset of agents that are capable to generate new blocks faster, imposing their version of the control history, as discussed in Section~\ref{sec:sec_issue}.

\subsection{Communication cost}

Fig.~\ref{fig:comparison} compares the communication exchanges in the blockchain-based scheme, Fig.~\ref{fig:comparison}(a), with the exchanges in the centralized one, Fig.~\ref{fig:comparison}(b). 
We proceed by evaluating and comparing the communication costs of both approaches, where the communication cost is expressed via the average amount of data exchanged per agent during a control period.

In the blockchain-based scheme, data is generated by the processes of smart contract updates and blocks generation.
The number of blocks generated during a control period, denoted by $N_b$, is determined by the protocol.
The communication cost for agent $u$ with $N$ peers is:
\begin{equation}\label{eq:bc_cost}
J^{bc} = N L_{u} + J^{rc} + p_b N_b N L_b + J^{rb},
\end{equation}
i.e., it comprises the costs of propagating its own transaction, expressed through the message length $L_u$, the relaying of other's transactions $J^{rc}$, the transmission of its own blocks and the relaying of other's blocks $J^{rb}$.
For the sake of comparison, we provide a lower bound on $J^{bc}$ for the simplified case where a DER has just a single peer and does not relay data\footnote{We note that the scenario with a single neighbor should be avoided, as in this case a malicious neighbor may corrupt or hide the information \cite{heilman2015eclipse}.}, assuming that the computational power is equally distributed among DERs, i.e., $p_b = 1 / U_{\text{total}}$:
\begin{equation}
J^{bc}_\text{LB} = L_u + \frac{N_b L_b}{U_{\text{total}}}.
\end{equation}

\begin{figure}[!tb]
\centering
  \includegraphics[width=0.92\columnwidth]{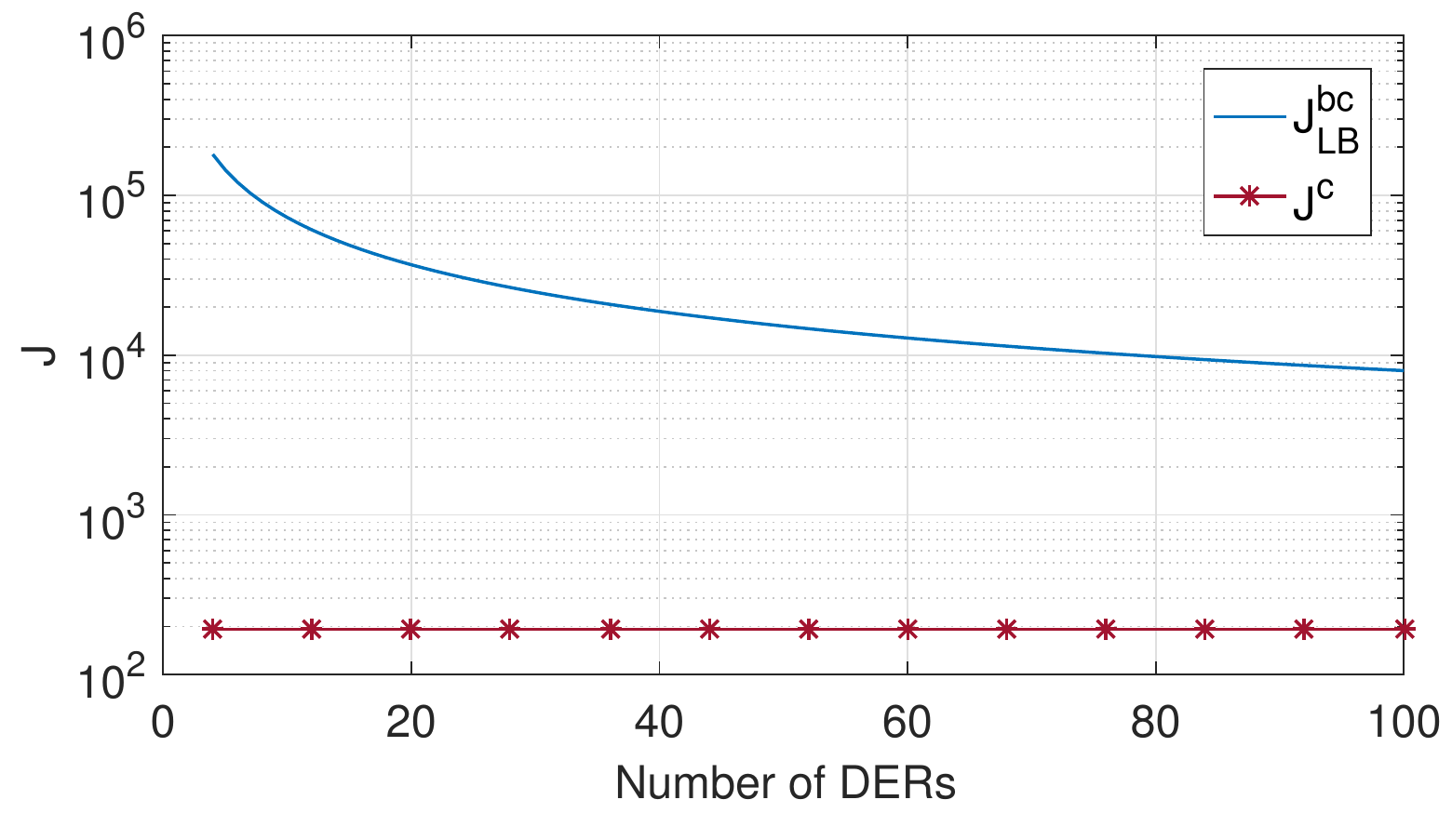}
    \caption{Comparison of communication costs.}
    \label{fig:commcost}
\end{figure}

In the centralized architecture, we denote the length of messages $w_u$, $c_u$, $\mu_u$ in bits as $L_{w}$, $L_c$, $L_{\mu}$.
In this case, the communication cost per DER is equal to $J^{c} = L_{w} + L_{c} + L_{\mu}$, i.e., it is a constant.  We compare the communication costs in Fig.~\ref{fig:commcost}, assuming that $L_w$, $L_{\mu}$, $L_c$ are 64 bits long, $L_u$ is 800 bits and $L_b$ is 8000 bits.
The block period is set to $10 \, \text{s}$, which together with $T^{tc} = 15 \, \, \text{minute}$  provides $N_b = 90$ blocks for control period.
The communication cost in the blockchain-based scheme is slightly decreasing with the number of DERs, i.e., $U_{\text{total}}$, as this reduces the number of blocks that they individually generate in a control period.
However, the centralized solution is clearly less demanding due to its simpler communication architecture.


\begin{figure}[!tb]
\centering

\subfloat[Without control.]{\includegraphics[width=0.88\columnwidth]{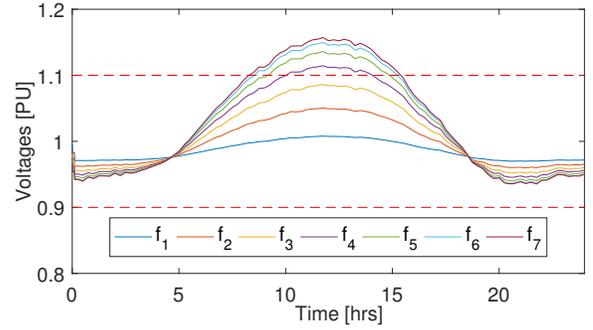}}

\subfloat[With the control.]{\includegraphics[width=0.88\columnwidth]{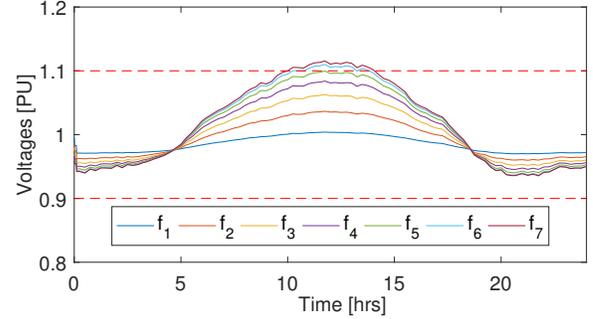}}
\caption{The voltage measured on the LV feeders on the dataset corresponding to 9th of June, indicated in the Per-Unit (PU) base.}\label{fig:voltage_day}
\end{figure}

\section{Case Study}\label{sec:case_study}

Inspired by the MG LV scenario in \cite{pedersen2015disc}, we adopt the power system composed by LV and MV MGs depicted in Fig.~\ref{fig:grid} as a case study. 
The LV MG is connected to a medium voltage (MV) microgrid in which a $6 \, \text{MW}$ solar PV power plant is installed.
The power exchange between the LV and MV MGs takes place via a on-load tap changing transformer (OLTC), that constitutes the point of common coupling (PCC).
The LV MG is composed by $N_{MG}=7$ feeders, each supporting the consumption of 10 households.
On each feeder, except feeder $f_1$, there are installed 4 PV systems (i.e., DERs) with the rated power of $4 \, \text{kW}$, resulting with a total of $U_{\text{total}} =24$ PVs.
For the modeling of PVs, OLTC, households consumption, and the other grid components, we adopt Disc framework \cite{pedersen2015disc}.

\begin{figure}[!tb]
\centering
  \includegraphics[width=\columnwidth]{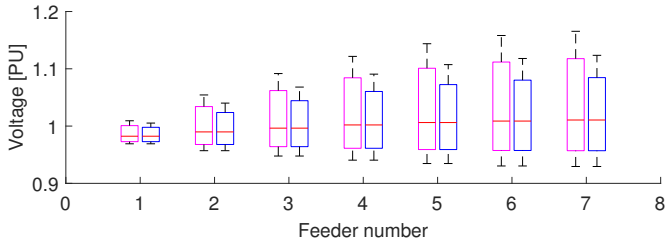}
    \caption{The voltage measured on the LV feeders over the month of June of Disc simulator (i) without control, depicted with purple boxes, and (ii) with proposed control, blue boxes. The plot shows minimum, quartile and maximum values.}
    \label{fig:month}
\end{figure}

We simulated the system in the Disc framework both for the case without control and the case in which the proposed control scheme is applied.
When no control is adopted for the PVs, in the afternoon hours we observe an overvoltage on the feeders, which increases with their distance from the transformer, see Fig.~\ref{fig:voltage_day}(a).
Fig.~\ref{fig:voltage_day}(b) shows the proposed control strategy\footnote{We used the optimized value of the droop parameter $\gamma = 0.005$, obtained via simulations.} provides a reduction of the over-voltages with respect to the scenario without control.
One can also observe that the slope of the voltage profile in afternoon/morning hours is reduced, which is another benefit of the control strategy.
Specifically, smoothing the voltage profile curve in systems dominated by PV power gives more time and flexibility to the bulk generation, which is characterized with high inertia and slow transient ramp-up/ramp-down response, to respond to the the power supply variations.
The proposed control strategy was also simulated over a period of one month and the reduction of over voltages was also verified, see Fig.~\ref{fig:month}.
Observe that the under-voltages are not reduced, as their control is not included in the proposed strategy.
Nevertheless, we note that a similar credit system can be employed for the load prioritization during low production periods.

The verification of the proportional fairness objective \eqref{eq:objective} in a centralized setting was performed via MATLAB simulations, where the initial credit was randomly distributed among DERs, i.e., the credit status vector $\mathbf{c}$ was randomly initialized, and the control period duration was set $T^{tc}=15$ minutes.
We performed 1000 simulation runs, corresponding to 250 hours of LVDG operation, and obtained that the DERs spend the equal fraction of time operating as VSCs (the detailed presentation of these results is omitted due to space constraints).

\begin{figure}[!tb]
\centering
  \includegraphics[width=\columnwidth]{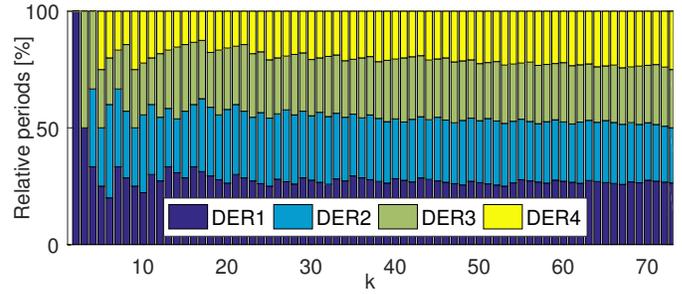}
    \caption{Evolution of the fraction of time spent by each DER as VSC in a feeder with 4 DERs.}
    \label{fig:testrpc}
\end{figure}

Finally, we turn to the blockchain-related aspects of the proposed framework.
The blockchain software installed in the DERs controller supported a private Ethereum blockchain \cite{wood2014ethereum}, which provides the possibility of writing complex smart contracts, and is simulated using EthereumJS testrpc \cite{testrpc}.
We deployed $N_{MG}$ contracts on the blockchain, where DERs interacted with the one corresponding to their feeder, and the initial credit was randomly distributed among DERs.
The scripts that interface DERs with the blockchain implemented the functionalities described in Section~\ref{sec:solution}.
We monitored the output of the scripts and observed that the contract was effectively reproducing the fairness objective, see Fig.~\ref{fig:testrpc}.
Clearly, after a transient period caused by the initial credit distribution, DERs operate as VSCs for equal fractions of time.

\section{Conclusions}\label{sec:conclusion}

The current research efforts in MG control are oriented towards distributed schemes, requiring development of novel protocols to enforce the security and the information trustworthiness among control agents.
The blockchain protocol has interesting properties that can be used to this end, resulting in a novel design of multi-agent control systems.
In this paper, we developed proportional fairness MG control and established a comparison between the standard centralized architecture and blockchain-based one, verifying that the blokchain-based solution can reproduce the control objectives of the centralized architecture.
We note that further investigation should be conducted with respect to more complex control schemes.

We also outline two potential limits of the private blockchain architecture in the context of MG control: the mining cost and the communication cost.
Specifically, alternatives to the energy inefficient POW have to be found in order to enable private blockchains for systems with limited hardware capabilities, such as MG components.
Secondly, the communication cost of the blockchain protocol is significantly higher than the cost of the centralized one.
The design of blockchain protocol tailored for MG applications is part of our ongoing research.

\section*{Acknowledgment}

The work presented in this paper was supported in part by the EU, under grant agreement no. 607774 ``ADVANTAGE".

\nocite{*}
\bibliographystyle{IEEEtran}
\bibliography{refs}



%

\end{document}